\documentstyle[aps,prb,eqsecnum]{revtex} 
\input epsf
\begin{document} \title{Fermi Edge Singularities: Bound-states and
Finite Size Effects} \author{Alexandre M. Zagoskin$^1$ and Ian
Affleck$^{1,2}$}

\address{$^1$Department of Physics and Astronomy and $^2$Canadian
Institute for Advanced Research,  The University of British Columbia,
 Vancouver, B.C., V6T 1Z1, Canada}  \maketitle
\abstract{Fermi edge adsorption singularities (FES) are studied using a
combination of conformal field theory (CFT), an exact sum rule and
numerical work on a tight binding model which is shown to exhibit
remarkable simplifying features. The  relationship between  FES and
the Anderson orthogonality exponent is established in great
generality, using CFT, including the case where the core hole
potential produces a boundstate.  Universal results on the adsorption
intensity in a finite sized sample are obtained.  Various predictions
 are checked numerically and the evolution of the
adsorption intensity with electron density is studied.}

 \section{Introduction}
A theoretical understanding of the Fermi edge singularity (FES) in
X-ray adsorption in metals and of the related Anderson orthogonality
catastrophe dates back to the
1960's.
\cite{Mahan
1,NdD,SS,Hopfield,Langreth,CN,Pardee-Mahan,Mahan 2,Girvin,Anderson} 
Nonetheless, this remains an active area of research today, in part
because of theoretical and experimental work in systems of reduced
dimensionality where strong correlation effects may play an important
role.\cite{Haw,AL,Fabrizio,Eder,ML,Geim,Shields,Young} In particular a
new and very general theoretical approach has been developed, based on
conformal field theory.\cite{AL} 

Recent experiments
have studied optical absorption threshold singularities in modulated,
doped semi-conductor layered structures.\cite{Shields,Young}  These behave more
or less like two dimensional metals with a continuously tunable
electron density.  The  potential produced by the valence band hole is
expected to produce a boundstate (exciton) due to the well-known
theorem that an attractive potential in two dimensions always has a
boundstate.  Many interesting
 issues are raised by these experiments which have not yet been
adequately addressed theoretically.  In particular, it is possible to
study the behaviour of the threshold singularities as the electron
density goes to 0.  In some cases an additional threshold
corresponding to a negatively charged exciton (two conduction
electrons bound to a valence hole) is observed.  Strong
correlation effects may play a crucial role in this case and the usual
Fermi liquid approach may need to be modified.  

The present work addresses several issues in this field, within the
usual Fermi liquid framework.  The case of a core potential producing a
boundstate is considered, as is the behaviour of the threshold
singularities as a function of electron density. Furthermore, we
consider the nature of the adsorption intensity for a  simple model of
a finite sample.  The techniques employed in this paper are a
combination of conformal field theory methods, an exact sum rule and
numerical work on a tight binding model which exhibits remarkable
simplifying features making it feasible to study very large systems.

The conformal field theory method of Ref. (\onlinecite{AL}) is extended
to the case where there is a boundstate.  A very simple  and
general proof of the exact correspondence between the FES exponent and
the orthogonality exponent is given.  The adsorption intensity
near threshold, for a finite system is shown to have a simple universal
form, using conformal field theory.  An exact sum rule is introduced
which determines the ratio of adsorption intensities for the two
cases where the core bound state is empty or filled.  

Numerical work on the simple tight binding model is used to
check the validity of Anderson's formula relating the
orthogonality exponent to the phase shift at the Fermi surface, the
Nozi\`eres-de Dominicis, Combescot-Nozi\`eres formulas\cite{NdD,CN} for
FES exponents including the boundstate case, and the formulas newly
derived here for the adsorption intensity in a finite system.  In
addition this numerical work, together with the sum rule, is used to
study the behaviour of the adoption intensity as a function of
density.

Some of the new conformal field theory results were briefly described
in Ref. (\onlinecite{A96}).



\section{Conformal field theory approach} An approach based on
boundary conformal field theory \cite{AL,A96} provides a unified  view
of the problem.  As usually, we start from the simplest possible model
\cite{NdD},  \begin{equation} {\cal H} = \sum_{\bf k}\epsilon_{\bf k}
c^{\dagger}_{\bf k}c_{\bf k} + b^{\dagger}b \sum_{{\bf k},{\bf k}'} V_{{\bf k},{\bf
k}'}c^{\dagger}_{\bf k}c_{{\bf k}'} + E_0 b^{\dagger}b.  \label{Eq.H} \end{equation}
Here the band of spinless, non-interacting electrons $(c^{\dagger},c)$ is
scattered by the core hole potential $V_{{\bf k},{\bf k}'}$. The core
hole $(b^{\dagger},b)$ is dispersion-less. It is created and annihilated
instantaneously (by irradiation), and the reaction of the band
electrons on this instantaneous perturbation constitutes the FES and
Anderson orthogonality effects.
 
In the low-energy limit,  the system is mapped onto the
$(1+1)$-dimensional Dirac Fermions defined on a ray $r>0$ with a
scattering potential $V$ at the origin. (This can be done by assuming
spherical symmetry of $\epsilon_k$ and $V_{k,k'}$ and considering only
s-wave scattering;  generalizations  to other cases are
straightforward.)  The following discussion  applies, with minor
modifications, to either an s-wave projected 3D (3 dimensional)
problem or to a problem defined a priori in 1D. 
 We will henceforth generally consider the 1D case.  We could
consider, for example, a 1D tight-binding model, defined on the
positive half-line, with free boundary conditions and a potential
localized near $x=0$.  See Appendix A for a detailed discussion of
this model.  The corresponding boundary condition in the low energy
Dirac theory is: \begin{equation} \psi_L(0)=\psi_R(0).\end{equation}
The role of the scattering potential in the theory here is to  impose
an effective boundary condition on the low energy degrees of freedom,
relating the left and right movers: \begin{equation} \psi_R(0) =
e^{2i\delta(k_F)}\psi_L(0),\label{bc} \end{equation}  Here $\delta
(k_F)$ is the phase shift at the Fermi surface; $k$-dependence of the
actual phase shift is irrelevant at low energies.
 The action of the hole creation operator, $b^\dagger,$  thus reduces
to that of a primary boundary condition changing operator, $\cal O$.  
The Green's function  (hole propagator) of this operator in a
half-plane, $z = r+ i\tau, r\geq 0$, is  \begin{equation}
G(\tau_1-\tau_2) \equiv \left<b(\tau_1)b^\dagger (\tau_2)\right> =
\left<A;0\right|{\cal O}(\tau_1){\cal O}(\tau_2)\left|A;0\right> =
\frac{1}{(\tau_1-\tau_2)^{2x}}. \end{equation} Here $x$ is the scaling
dimension of $\cal O$, and $\left|A;0\right>$ is the
 ground state of the infinite system (filled Fermi sea) without
scattering potential.  Physically, the Green's function is directly
related to the 
 absorption intensity in the case of photoemission,  \begin{equation}
I(\omega) \propto \int dt e^{i(\omega -\omega_0)t}
\left<b(t)b^{\dagger}(0)\right>  \propto (\omega - \omega_0)^{-\alpha},
\label{Eq.0} \end{equation} where $\omega_0$ is the threshold
frequency. Evidently, the (FES-) exponent $\alpha$ and the scaling
dimension are related via \begin{equation}
 \gamma \equiv 1-\alpha  = 2x. \end{equation}

Looking for  finite size effects, we conformally map the half-plane 
onto the strip, $l\geq r\geq 0$ Using the transformation $z = l e^{\pi
w/l}$.  In a bosonic system this automatically gives the same boundary
condition at $0$ and $l$.  However, for fermions it gives:
\begin{equation}\psi_L(0)=\psi_R(0),\ \  \psi_L(l)=-\psi_R(l).
\label{bcs} \end{equation} This follows because the fermion fields
transform as: \begin{eqnarray}\psi_L&\to& (dz/dw)^{1/2}\psi_L\nonumber
\\ \psi_R&\to& (dz^*/dw^*)^{1/2}\psi_R.\end{eqnarray}  At $w=x+il$, 
\begin{equation}(dz/dw)^{1/2}/(dz^*/dw^*)^{1/2}=-1.\end{equation} This
transformed problem corresponds to considering a 1D model defined on a
finite line, $0<x<l$, with the impurity potential near $x=0$ and an
appropriate boundary condition at $x=l$.  Alternatively, the 3D s-wave
projected system is now defined inside a finite sphere of radius $l$
with an appropriate boundary condition on the surface of the sphere.
For a discussion of this boundary condition and more details, see
Appendix A.

We find for the Green's function on the strip \begin{equation}
\left<AA;0\right|{\cal O}(u_1){\cal O}(u_2)\left|AA;0\right> =
\frac{1}{\left(\frac{2l}{\pi}\sinh \frac{\pi}{2
l}(u_1-u_2)\right)^{2x}}, \label{GFstrip} \end{equation}
  $\left|AA;0\right>$ being the unperturbed ground state of the system
of length $l$, with the ``same'' boundary condition, $A$, given by Eq.
(\ref{bc}) at both ends.  

In Eq. (\ref{GFstrip}) we can either Taylor expand $\sinh$ in the limit
$\pi(u_1-u_2)/l \ll 1$, or insert a complete set of states
$\left|AB;m\right>$ (eigenstates of the system with the scattering
potential - boundary condition ``B'' - present), and obtain the
relation \begin{eqnarray}
 && \left(\frac{\pi}{l}\right)^{2x} e^{-\frac{\pi x (u_1-u_2)}{l}}
\Bigl( 1 + 2x e^{-\frac{\pi(u_1-u_2)}{l}} + \frac{2x(2x+1)}{2}
e^{-\frac{2\pi(u_1-u_2)}{l}}  \nonumber \\ &&
+ \frac{2x(2x+1)(2x+2)}{6}
e^{-\frac{3\pi(u_1-u_2)}{l}}
+\dots \Bigr)   =\sum_m
\left|\left<AA;0\right|{\cal O}\left|AB;m\right>\right|^2
e^{-[E_m^{AB}- E_0^{AA}][u_1-u_2]}.\label{Taylor} \end{eqnarray}

If for the operator $\cal O$ the first nonvanishing matrix element is
with the ground state of the perturbed system, $\left|AB;0\right>$, 
  then for the overlap of the two ground states  (Anderson
orthogonality catastrophe) \begin{equation}
\left|\left<AA;0\right|{\cal O}\left|AB;0\right>\right| =
\left(\frac{\pi}{l}\right)^{x}. \label{aoe}\end{equation} The 
Anderson orthogonality exponent coincides with the scaling  dimension
$x$; on the other hand, $x$ is given by the O$(1/l)$- contribution to
the ground state energy shift due to the perturbation,\cite{AL,A96}
\begin{equation} x=\frac{l}{\pi}[E^{AB}_0 - E^{AA}_0]. \end{equation}
Here it is being implicitly assumed that this energy difference
consists of a term of $O(1/l)$ only, as would follow from conformal
invariance.  As discussed in Appendix A there will in general also be
a term of $O(1)$ which must be subtracted.

 A simple way of determining the Fermi edge exponent $\gamma$ and the
orthogonality exponent $x$, is thus to calculate the $1/l$ finite-size
correction to the difference in groundstate energies of the system
with and without the scattering potential.   The term of O(1) is
non-universal, (cut-off dependent) while the higher order terms
contain the corrections from various irrelevant operators.  On the
other hand, we expect the term of $O(1/l)$, which is determined only
by the immediate vicinity of the Fermi surface, to be universal and to
give the desired FES and orthogonality exponents.  In fact, this
result remains true including bulk Coulomb interactions in one dimension.\cite{AL} 
Thus calculation of the $O(1/l)$ term in the groundstate energy
difference gives a very simple way of determining the FES and
orthogonality exponents in great generality.  This calculation is
spelled out in detail, for a one-dimensional tight-binding model, in
App. A.  The conclusion is: \begin{equation}x={1\over 2}\left[\delta
(k_F)\over \pi \right]^2.\end{equation}

While various derivations of this result, both for the FES exponent
and for the orthogonality exponent have been given before, this one
has certain distinct advantages.  The original derivation of the
orthogonality exponent by Anderson made a variety of approximations,
including Taylor expanding certain quantities in powers of the phase
shift.  The derivation of the FES exponent in[\onlinecite{NdD}] also
initially assumed a small $\delta (k_F)$ and then argued for the
generality of the result by some fairly subtle consistency arguments. 
The bosonization derivations start with a bosonized Hamiltonian
written in terms of $\delta (k_F)$ whereas a naive bosonization in
fact only picks up the Born approximation to $\delta$, linear in the
scattering potential. It is expected that eliminating the high energy
modes somehow renormalizes this parameter in the bosonized Hamiltonian,
turning it into the true phase shift.  Once the assumption of
conformal invariance is made, it is very straightforward to
demonstrate that it is precisely the phase shift at the Fermi surface
which enters the exponents, by an explicit calculation of the
groundstate energy, as given in App. A. We note that once the
bosonized Hamiltonian is assumed, the results for the strip can be
obtained by a  mode expansion of the boson field.  This, of course,
gives the same result obtained more simply by the conformal
transformation.

Another advantage of this somewhat abstract approach to the problem is
that it can be immediately generalized to the case where the core
potential creates a boundstate. The Green's function can then be
presented as a sum of two terms:
 \begin{eqnarray}
   G(u) &=& G_e(u) + G_f(u)  = \sum_m
\left|\left<AA;0\right|{\cal O}\left|AB;m;e\right>\right|^2
e^{-[E_{m,e}^{AB}- E_0^{AA}]u} \nonumber \\
&&+ \sum_n \left|\left<AA;0\right|{\cal
O}\left|AB;n;f\right>\right|^2 e^{-[E_{n,f}^{AB}- E_0^{AA}]u},
\label{Green}\end{eqnarray} where the first sum is taken over all
states $\left|AB;m;e\right>$ where the  boundstate is empty, and the
second over the states  $\left|AB;n;f\right>$ where it is filled.
These two terms give rise to two peaks in the absorption  rate,
separated by the binding energy, $\Delta\omega = |E_B|$.  Introducing
the occupation number of the boundstate, $\hat n_B$,  \begin{equation}
G(u) = <[b(u)\hat n_B(u)][\hat n_B(0)b^\dagger (0)]> +<[b(u)(1-\hat
n_B(u))][\hat (1-n_B(0))b^\dagger (0)]>.\end{equation} In the long
time limit we may calculate each of these terms separately using the
boundary conformal field theory approach.  The boundstate is
associated with a finite binding energy, $E_B$ and an exponentially
decaying wave-function.  Thus it has no direct effect on the $O(1/l)$
terms in the energies.  Therefore we expect the above formulas to
apply immediately for the first threshold where the boundstate is
filled, with the exponent: \begin{equation} x_f={1\over
2}\left[{\delta (\epsilon_F)\over \pi}\right]^2.
\label{x_f}\end{equation}

When the boundstate is empty, the only change in the low energy
physics is that one additional electron is raised to the first
unoccupied state above the Fermi surface. 
 This  has wave-vector: \begin{equation} k = k_F+{\pi \over
l}\left[{1\over 2} -\delta (\epsilon_F)\right]. \end{equation} We may
regard $b[1-n_B]$ as a different boundary condition changing operator
which creates one additional low energy electron, in addition to
producing the new boundary condition of Eq. (\ref{bc}).  The $O(1/l)$
term in the ``groundstate'' energy difference in the case of the empty
boundstate is, from App. A: \begin{equation}  E_0'-E_0=v_F{\pi \over
l}{1\over 2}\left[{\delta (\epsilon_F)\over \pi}-1\right]^2
\end{equation}   Thus the orthogonality exponent giving the overlap
between the unperturbed groundstate and the ``groundstate'' with the
boundstate empty is: \begin{equation} x_e={1\over 2}\left[{\delta
(\epsilon_F)\over \pi}-1\right]^2. \label{x_e}\end{equation}

The FES exponents for the two thresholds $\alpha_f$ and $\alpha_e$ are
given by: \begin{equation} \alpha_f=1-2x_f,\ \ \ 
\alpha_e=1-2x_e.\label{alpha}\end{equation}  By merely computing the groundstate
energy, rather than attempting to compute the exponents directly, we
have finessed the problem of attempting to bosonize the theory with
the boundstate. We note that these results agree with Combescot and
Nozi\`eres\cite{CN} and Hopfield \cite{Hopfield}.  The present
derivation seems quite closely related to the observation of Hopfield
that the FES exponent measures the amount of charge pulled in from
infinity by the boundstate. 


The conformal mapping from the plane to the strip establishes in a
simple way the relationship between the FES exponents of the infinite
system and the orthogonality exponents and energies of the finite
system.  In fact this mapping provides considerably more information. 
Let's imagine a rather artificial situation where a core hole is
instantaneously created at the end of a finite one-dimensional
system.  (Equivalently we could consider an artificial situation where
it is created at the centre of a finite sphere.)  In this case the
adsorption intensity, $I(\omega )$ of Eq. (\ref{Eq.0}) becomes a
series of $\delta$-function peaks, as we see from Fourier transforming
Eq. (\ref{Taylor}). The peaks occur at $\omega_m = E_m^{AB}- E_0^{AA}
= \pi(x+m)/l $, the energies of excited states of the perturbed system
measured from the unperturbed ground state energy $E_0^{AA}$. [The
neglected term of $O(1)$ just shifts the threshold by $\omega_0$.]
 It follows from (\ref{Taylor}) that the ratio of the $m$th peak to
the zeroth one depends only on $x$: \begin{equation}
 \frac{\left|\left<AA;0\right|{\cal O}\left|AB;m\right>\right|^2}{
\left|\left<AA;0\right|{\cal O}\left|AB;0\right>\right|^2}  = 
\frac{2x(2x+1)(2x+2)\dots(2x+m-1)}{m!}. \label{peaks} \end{equation}
Note that, from Eq. (\ref{Taylor}), each of these peak intensities
scales with length the same way as does the $0^{\hbox{th}}$ peak,
considered by Anderson, given by Eq. (\ref{aoe}).  The ratios of peak
intensities are length independent pure universal numbers, determined
only by $\delta (k_F)$.

 In fact, each of these peaks corresponds, in general, to several
different states with energies that are degenerate, to $O(1/l)$. 
These are simply multiple particle-hole excitations of the free
fermion system, with a dispersion relation which is linearized and a
phase shift which is assumed $k$-independent, near the Fermi surface. 
A general particle-hole excitation may be constructed by first raising
$n_m$ electrons $m$ levels, then raising $n_{m-1}$ electrons $m-1$
levels, etc.  The energy relative to the groundstate of this state is: 
\begin{equation} E_m^{AB}-E_0^{AB} = {\pi \over l}\sum_{p=1}^\infty
n_pp.\end{equation}The first excited state has $n_1=1$, $n_p=0$ for
$p>1$.  The next degenerate pair of states have $n_1=2$ or $n_2=1$
(with the other $n_p=0$).  The third set of excited states is
threefold degenerate with $n_1=3$ or $n_1=n_2=1$ or $n_3=1$ (and the
other $n_p=0$ in all cases).  Corrections to the linear dispersion
relation and variation of $\delta (k)$ near $k_F$ will split these
energies by amounts of $O(1/l^2)$.  The simple prediction obtained
here from a conformal transformation does not give the amplitude of
each peak separately, but only  the sum of amplitudes of all peaks at
a given energy, where energy differences of $O(1/l^2)$ are ignored.  

This new finite size result interpolates, in a sense, between the
orthogonality exponent and the FES exponent.  Considering the large
$m$ limit of Eq. (\ref{peaks}) we find that the intensity decays as
$m^{-(1-2x)}$, recovering the FES exponent.

This result applies immediately to the peaks corresponding to the
boundstate being
 filled or empty, provided that the appropriate orthogonality
exponents, $x_f$ and $x_e$ of Eq. (\ref{x_f}) and (\ref{x_e}) are used.

\section{One-dimensional tight-binding model}

An evident discrete counterpart to the system (\ref{Eq.H}) in its one-
dimensional version is the system of spinless fermions  on a
finite 1D chain with nearest-neighbour hopping, free boundary
conditions and  an impurity potential which can be switched on/off at
the first site:
 
\begin{eqnarray} {\cal H} = H_0 + b^{\dagger}b H_1;\nonumber \\
 H_0 = -t\sum_{i=1}^{l-2}
\left(\psi^{\dagger}_i\psi_{i+1} + \psi^{\dagger}_{i+1}\psi_i \right);
\nonumber \\ 
H_1 = -V \psi^{\dagger}_1\psi_1.
\label{Hsolv}\end{eqnarray}
 Here we choose $V$ to be positive in case of an attractive
core potential, and also choose $t>0$.
This model is very amenable to large scale numerical work with a
minimum of effort.  Not only can the single particle energies and
wave-functions for finite $l$ be found exactly in a simple form, but,
more remarkably, the overlaps of the single particle wave-functions
corresponding to different values of the potential, $V$, obey an exact
factorization.  This enormously simplifies the calculation of the
overlap of the many-particle, Bloch determinant, wavefunctions.  This
can then be expressed by the Cauchy determinant formula, used as an
approximation by Anderson in his classic paper on the orthogonality
catastrophe.\cite{Anderson}  The many-body overlaps are easily evaluated
numerically and in some limits analytically (e.g. in the narrow band
limit, $\frac{t}{V}\to 0$). As a result,  chains of length up to few
thousand sites are easily handled on a workstation using the standard
``Mathematica'' package. The crucial factorizability of
one-particle overlaps disappears under any other   position of the
scattering potential.  

\begin{figure}[p]
\epsfysize = 3 in 
\epsfbox{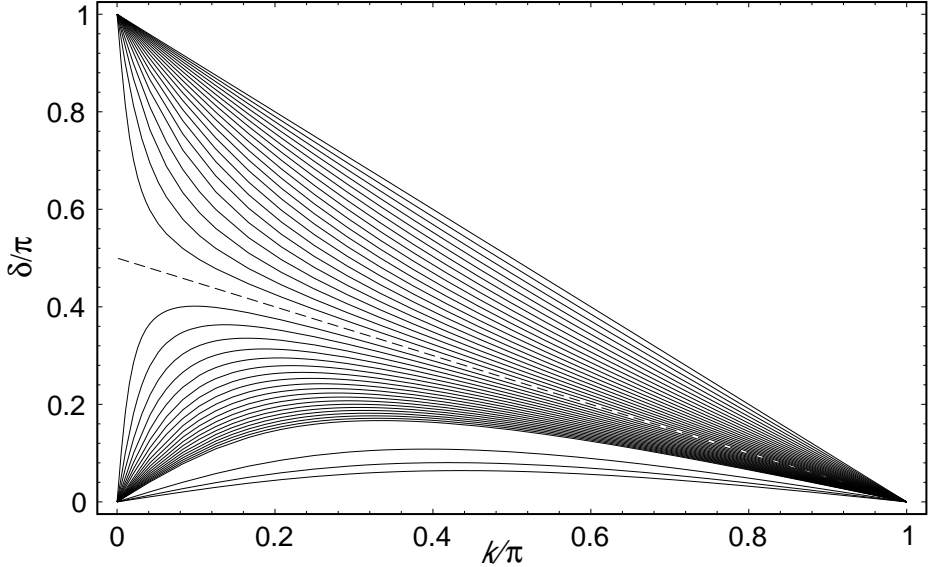} 
\caption{Phase shift in the 1D
tight-binding model vs. density $\nu = \lim_{l\to\infty} N/(l-1) $,
at $t/V =0,0.05.0.01,\dots,2,3,4,5$.
 The dotted line $t/V=1$ separates
regions with and without boundstate.}
\label{Figure:delta} \end{figure}

The details are given in the Appendix B; here we just summarize a few
salient features, beginning with the infinite $l$ limit.  This model has a
band of eigenstates with exact wave-functions: \begin{equation} \Psi_j
\propto \sin [kj+\delta (k)], \ \  j=1,2,3,\ldots \end{equation} The
dispersion relation is: \begin{equation} \epsilon (k)=-2t\cos
k.\end{equation}  The phase shift is given by: 
\begin{equation}\delta =
\arctan\left[{\sin k\over t/V-\cos k} \right].
\end{equation}
 Note that
at the bottom of the band, $k\to 0$, $\delta \to 0$ for $V<t$, when there
is no boundstate, but $\delta \to \pi$, for $V>t$ when there is a
boundstate, as required by Levinson's theorem.  As $k$ ranges over the
whole band, from $0$ to $\pi$,  $0\leq \delta (k)\leq \pi /2$ for $V<t$,
and $0 \leq \delta (k)\leq \pi$  for $V>t$, as shown in Fig.
(\ref{Figure:delta}). For $V \gg t$, $\delta(k) \approx \pi-k.$ There is one boundstate, if $V>t$ only, with:
\begin{equation}E_B=-(V+t^2/V).\end{equation} Note that this approaches
the bottom of the band, $-2t$, as $V\to t$,
where the boundstate disappears.  
The exact boundstate wavefunction is:
\begin{equation} \Psi^B_j \propto e^{-\kappa j},\end{equation} with
\begin{equation} \kappa = \ln (V/t).\end{equation}

For finite $l$, there is a set of wavefunctions:
 \begin{equation}\Psi^n_j\propto  \sin \tilde k_n(j-l),\end{equation}
 with the allowed
wave-vectors determined by: \begin{equation} {\sin k(l-1)\over \sin
kl}={t\over V}.\label{detk3} \end{equation}
For $t/V>1-1/l$ there are $l-1$ solutions of Eq. (\ref{detk3}), which we
label \begin{equation} \tilde k_1,\tilde k_2,\ldots \tilde
k_{l-1}.\end{equation} 
For $t/V<1-1/l$ there are only  $l-2$ such wavefuctions, which we label
$2, 3, \ldots (l-1)$ and an additional wavefunction:
\begin{equation} \tilde \Psi^1_j\propto \sinh \kappa
(j-l),\end{equation} with $\kappa$ the solution of:
 \begin{equation} {\sinh \kappa (l-1)\over \sinh
\kappa l}={t\over V}.\label{kappa3}
\end{equation}

In the case $V=0$, the wave-vectors are:
\begin{equation} k_n=\pi n/l,\ \  n=1,2,3,\ldots (l-1). \end{equation}
From Eq. (\ref{detk3}) we see that, in the limit $V/t\to \infty$, the
wave-vectors are:
\begin{equation} \tilde k_n=(n-1)\pi /(l-1),\ \  n=2,3,4,\ldots (l-1),
\end{equation} corresponding to a chain of $(l-2)$ sites and a free
boundary condition.  The lowest wavefunction (boundstate) becomes
localized at $j=1$ in this limit with eigenvalue $-V$.  

The overlaps between the unperturbed, $ \psi$, and
perturbed, $\tilde \psi$, one-particle states assume the special form,
reminiscent of the first-order perturbation theory, but actually exact
(Appendix B):
 \begin{eqnarray} <\tilde \Psi^m|\Psi^n> &=&   
-V\frac{C(\tilde k_m)C(k_n)}{\epsilon (\tilde k_m)- \epsilon (k_n)},
\label{overlap-band}\\ 
<\tilde \Psi^1|\Psi^n>&=& -V
\frac{C_BC(k_n)}{\epsilon_B - \epsilon (k_n)}, 
\label{overlap-bound}
\end{eqnarray} where 
\begin{eqnarray}
 C(k)& =& \frac{\sqrt{2}\sin (l-1)k}{\sqrt{(l-1)- \sin (l-1)k
\cos lk/\sin k}},\\
 C_B   &=& \frac{\sqrt{2}\sinh (l-1){\kappa}}{\sqrt{ \sinh
(l-1)\kappa \cosh l\kappa /\sinh \kappa 
-(l-1)}}. \end{eqnarray}
 
\section{Calculation of hole propagator. Comparison to the CFT
predictions}

The Green's function $G(u)$ is determined by the set of matrix
elements, \begin{eqnarray} \left|\left<AB;m\right|{\cal
O}\left|AA;0\right>\right|  \to 
\left<\tilde{\Phi}_{m;e,f}\right|\left.\Phi_0\right> \equiv
C^{0m}_{e,f} \end{eqnarray} between (perturbed and unperturbed)
many-body states of the system and corresponding excitation energies,
$\Delta \epsilon_{m;e,f}$ (we note explicitly whether the boundstate if
present is occupied ($f$) or empty ($e$), and label by $m'$ the
appropriate excited states of the band).
  
When there are $N$ spinless noninteracting electrons in the system,  
the many-particle wave function  is a $N \times N$ Slater determinant
\begin{eqnarray} \Phi =
\frac{1}{\sqrt{N!}}\det(\Psi_{l_a}^{(n_b)}).\end{eqnarray} Here $
\Psi_{l_a}^{(n_b)}$ is an appropriate one-particle eigenfunction of
the state $n_b$, taken at the coordinate of the $a$th particle,  $a,b
= 1,2,\dots N$.

The overlap of two such states is a determinant \begin{equation}
(\tilde{\Phi},\Phi) = \frac{1}{N!}\sum_{j_1}\dots\sum_{j_N} 
\det(\tilde{\Psi}_{j_n}^{(m)*}) \det(\Psi_{j_n}^{(m)}) \equiv 
\det\left( (\tilde{\Psi}^{(m)},\Psi^{(n)}) \right).
\label{many_body_det} \end{equation}

  The remarkable form of the one-particle overlaps
(\ref{overlap-band},\ref{overlap-bound}) allows us to apply
the Cauchy formula \cite{Polya} in order to calculate the determinant
(\ref{many_body_det}):

 \begin{eqnarray} \det\left(\frac{1}{a_m+b_n}\right) = 
\frac{\prod_{m>n}(a_m-a_n)\prod_{m>n}(b_m-b_n)}{\prod_{m,n}(a_m+b_n)}
\end{eqnarray} Unlike the situation considered in the papers by
Anderson and Combescot and Nozi\'eres \cite{Anderson,CN},  in our model
the special form of the overlaps is an exact result, and not the
consequence of the linearization of the dispersion law close to Fermi
surface.

We begin with calculating $C^{00}_f$, the overlap of the  ground
states of the system with and without the core potential, which yields
the Anderson exponent (\ref{aoe}). In the corresponding
determinant (\ref{many_body_det})   both ``old'' and ``new''  indices
run  from $n,m=1$ to $n,m=N$. (That is, in the ``new'' state there is
one bound electron and $(N-1)$ electrons in the band, and no $e-h$ 
pairs). The other interesting overlap, $C^{00}_f$, corresponds to the
situation when the bound state in the ``new'' system is empty, and all
$N$ electrons are in the band, occupying the lowest lying states
(still no $e-h$ pairs):   $n=1,\dots,N$, but $m=2,\dots,N+1$. It
should yield the ``empty boundstate'' Anderson exponent, which
according to general considerations \cite{Hopfield} should be
$\left(1-\frac{\delta_F}{\pi}\right)^2$, as distinct from  the
``filled'' value  $\left(\frac{\delta_F}{\pi}\right)^2$. Using the
Cauchy formula, we find\begin{eqnarray} C^{00}_f = (-V)^N C_B
\prod_{m=2}^N C(\tilde{\kappa}_m)\prod_{n=1}^N C(\kappa_n)
\frac{\prod_{m>m'=1}^N(\tilde{\epsilon}_m-\tilde{\epsilon}_{m'})
\prod_{n>n'=1}^N(\epsilon_{n'}-\epsilon_n)} {\prod_{m=1}^N\prod_{n=1}^N
(\tilde{\epsilon}_m-\epsilon_n)}; \label{Cf} \\ C^{00}_e = (-V)^N 
\prod_{m=2}^{N+1} C(\tilde{\kappa}_m)\prod_{n=1}^N C(\kappa_n)
\frac{\prod_{m>m'=2}^{N+1}(\tilde{\epsilon}_m-\tilde{\epsilon}_{m'})
\prod_{n>n'=1}^N(\epsilon_{n'}-\epsilon_n)}
{\prod_{m=2}^{N+1}\prod_{n=1}^N
(\tilde{\epsilon}_m-\epsilon_n)},\label{Ce} \end{eqnarray} and for
their ratio: \begin{eqnarray} R = \frac{C^{00}_e}{C^{00}_f} = 
\frac{C(\tilde{\kappa}_{N+1})}{C_B} \cdot 
\frac{\epsilon_1-\tilde{\epsilon}_1}{\tilde{\epsilon}_{N+1}-\epsilon_1}
\cdot  \prod_{m=2}^N
\frac{\tilde{\epsilon}_{N+1}-\tilde{\epsilon}_m}{\tilde{\epsilon}_{N+1}
-\epsilon_m} \prod_{m=2}^N
\frac{\epsilon_m-\tilde{\epsilon}_1}{\tilde{\epsilon}_m-
\tilde{\epsilon}_1}.\label{Ratio} \end{eqnarray} 
Expression
(\ref{Ratio}) is  easily  calculated  for pretty large systems, since
it involves the number of operations only of order $l$. The approximate
values for the energies $\tilde{E}_m$    can be  accurately calculated
as a perturbation series in $t/V \ll 1$ (see Appendix B).  If
the above predictions are valid, then the ratio   should depend on
the system size as \begin{equation} R \propto
(l-1)^{\frac{1}{2}\left(\frac{\delta_F}{\pi}\right)^2 -
\frac{1}{2}\left(1-\frac{\delta_F}{\pi}\right)^2} =
(l-1)^{\frac{\delta_F}{\pi}-\frac{1}{2}}. \end{equation}

\begin{figure}[p]
 \epsfysize= 4 in \epsfbox{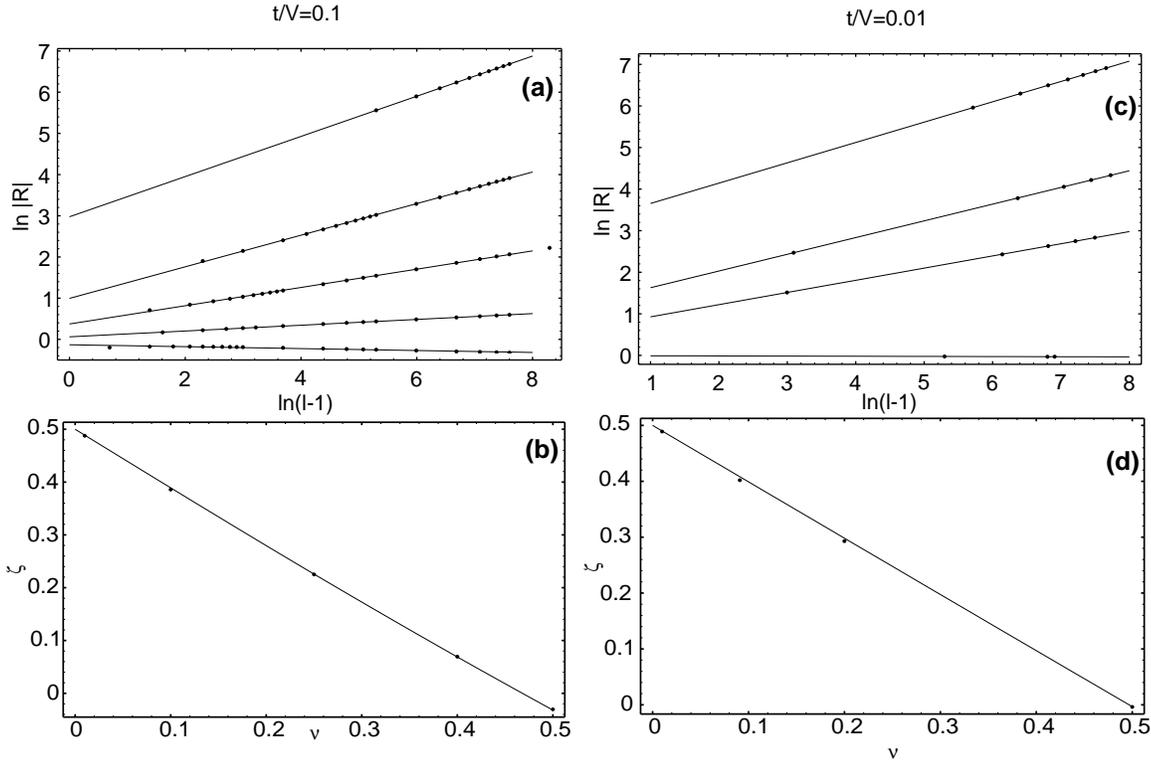}
 \caption{(a)Logarithm of ratio
$|R|=\left|C^{00}_e/C^{00}_f\right|$ as a function  of (even) chain
length $(l-1)$
for $t/V=0.1$ and $\nu = 1/2; 2/5; 1/4; 1/10$ and $1/100$; the
lines are best fits to points with $(l-1)>10$ by $\ln|R| = A + 
\zeta \ln(l-1)$; (b)
Exponent $\zeta(\nu)$. The curve is   $\frac{\delta(\nu;\frac{t}{V}=0.1)}{\pi}-\frac{1}{2}$.
(c)The same as (a) for $t/V=0.01$ and $\nu = 1/2; 1/5; 1/11$ and
$1/101$ (in the latter three cases the length $(l-1)$ was chosen to be
odd); the lines are best fits to points with $(l-1)>10$ by $\ln|R| = A
+ \zeta \ln(l-1)$; (d) Exponent $\zeta(\nu)$. The curve is
$\frac{\delta(\nu;\frac{t}{V}=0.01)}{\pi}-\frac{1}{2}$.} \label{Figure:R 0.1} \end{figure}
Here $\delta_F$ is the phase shift at the Fermi surface, $k_F = \pi\nu.$

The results are shown in Fig.\ref{Figure:R 0.1} ($t/V=0.1$  and $0.01$) for $(l-1)\leq 4000.$  As is clear from the figures, the
ratio as a function of $l$ at fixed density $\nu = \frac{N}{l-1}$
behaves as 

\begin{equation} R(l; \nu) \propto (l-1)^{\zeta(\nu)}, \end
{equation} where  indeed  (the least squares best fit parameters are
shown in figure captions): \begin{equation} \zeta(\nu) \approx
  \frac{\delta_F(\nu;\frac{t}{V}=0.01)}{\pi} - \frac{1}{2},
 \end{equation} 
for $\frac{\delta_F(\nu)}{\pi} \approx \frac{1}{2} - \nu,$ as seen in Fig.1.

\begin{figure}[t]
\epsfysize= 4 in \epsfbox{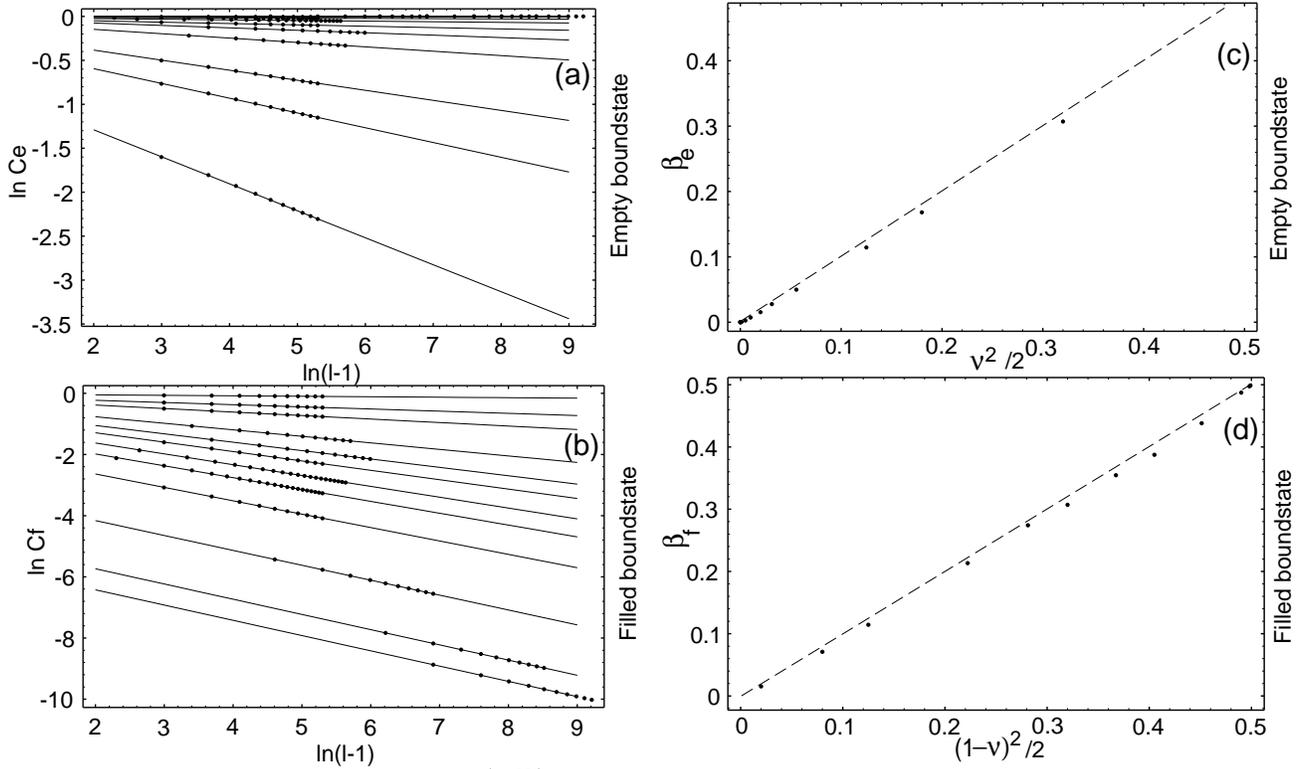}
\caption{(a)Logarithm of the coefficient $\left|C^{00}_e\right|$
(empty boundstate) as a function  of  chain length for $t/V\to 0$ and
$\nu = 1/1000;1/500;1/100;1/20; 1/7;1/5;1/4;1/3;1/2;3/5$ and $4/5$;
the lines are best fits to points with $(l-1)>10$ by $C^{00}_e(l,\nu) =
A_e(\nu) N^{-\beta_e(\nu)}$; (b) The same for $\left|C^{00}_f\right|$
(filled boundstate). (c,d) Anderson exponents $\beta_{e}$ (c)
and $\beta_{f}$ (d). The  curves are  $\beta_e(\nu)  =  \frac{\nu^2}{2}$ and $\beta_f(\nu) =  \frac{(1-\nu)^2}{2}.$ (In the limit $t/V=0$,  $\nu = 1-\delta/\pi$.)} \label{Fig:AndersonExp} \end{figure}

Calculation of the coefficients $C^{00}_{f,e}$ themselves is   more
time consuming. We calculated $C^{00}_e$ and $C^{00}_f$  in the limit
$t/V\to 0$ for $(l-1) \leq 400$. The results are shown  in
Fig.\ref{Fig:AndersonExp}. The matrix elements 
 dependence on size and density is accurately described by
\begin{equation} C^{00}_{e,f}(l,\nu) = A_{e,f}(\nu)
(l-1)^{-\frac{1}{2}\left(\frac{\tilde{\delta_F}(\nu)}{\pi}\right)^2},
\end{equation} thus  confirming the validity  of the original 
Anderson's result \cite{Anderson} in the case of binding core potential
and arbitrary  electronic density.

\begin{figure}[t]
\epsfysize= 3.7 in \epsfbox{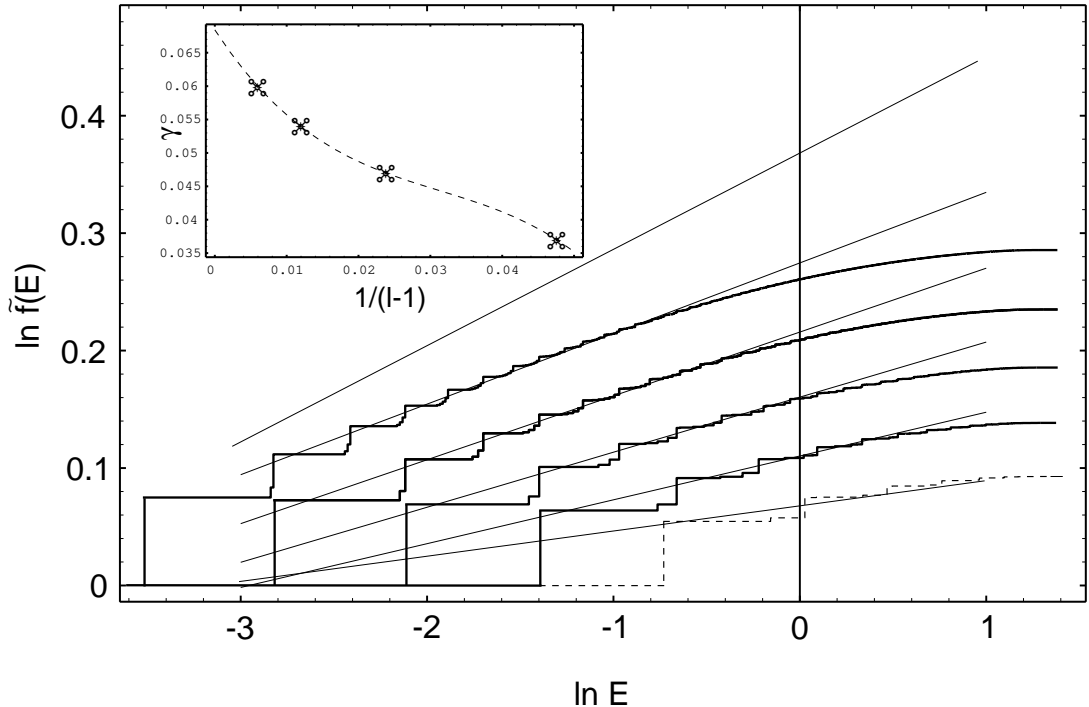}
\caption{The  normalized integral  of intensity,  $\tilde{f}(E) = \sum_{E_m\leq E}\frac{\left|C^{0m}_{e}\right|^2}{\left|C^{00}_{e}\right|^2} \propto \int_0^E
I(\omega) d\omega$, 
vs. energy, for the case of empty boundstate and $t/V\to 0$. Density
$\nu = 6/21;$ system size $(l-1)=21; 42; 84; 168$.  The best fits yield
$\gamma = 0.037;0.047;0.054$ and $0.060$ resp. The CFT value is
$0.082$ (uppermost line).
 The clusters of almost degenerate excited states are clearly
seen (in one-pair approximation the $n$-th excited state is
$n$-fold degenerate). The dotted line shows the direct diagonalization 
results by Eder and Sawatzky (Ref.(\protect{\onlinecite{Eder}})), for $t=1, V=32, N=6, (l-1)=21;
\gamma=0.021.$  Due to normalization to $\left|C^{00}_{e}\right|^2$ the
 curves are offset in vertical direction; as the system size grows,
the plateau heights are reaching  the universal values predicted by CFT for the ratios of subsequent excited
peak amplitudes to the lowest energy peak amplitude (see Eq.(\protect{\ref{peaks}}) and Fig.\protect{\ref{Figure7}}). 
Inset: FES exponent vs. inverse chain length. The cubic extrapolation to infinite
system size $\gamma(0) = 0.068$ still falls short of the CFT value.} \label{Figure:cumulative} \end{figure}

 Turning to the contributions of the excited states, $C^{0m'}_{f,e}$,
we must keep in mind that in the model, the degeneracy of energy
levels in
  the CFT formula (\ref{Green}) is lifted.
 Therefore when checking Eq. (\ref{peaks}), we  calculate the  ratios $
\frac{\left|\sum_m'
C^{0m}_{e,f}\right|^2}{\left|C^{00}_{e,f}\right|^2},$ where the sum is
extended over the excited states which would be degenerate in the case
of linear dispersion relation. These clusters of nearly degenerate
states are clearly seen in Fig. (\ref{Figure:cumulative}), where we plot
$\log f(E) = \log\sum_{\tilde \epsilon_m\leq
E}\left|C^{0m}_{e,f}\right|^2$ vs. $\log E$. 
In fact, not all excited
states are included in Fig. (\ref{Figure:cumulative}), but only those
corresponding to a single particle-hole excitation.  In the large $l$
limit,  all excited states with energies $\pi v_Fm/l$ with $m=1,2$
or $3$, discussed in Sec. II, correspond to single particle-hole
excitations.  However, multi particle-hole excitations begin to appear
at $m=4$.  We might try to extract the FES exponent from the slope at
low energies and large $l$, assuming that the neglected multi
particle-hole excitations don't make too large a contribution close to
the threshold.  Assuming $I \sim \omega^{-\alpha}$ in this region, then
for small $E$ $f(E) \sim  E^{1-\alpha} \equiv E^{\gamma}$. The tangent
of the curve should yield  the FES exponent-related $\gamma =
\left(\frac{\tilde{\delta}_F}{\pi}\right)^2$, but actually it is
significantly smaller.

\begin{figure}[t]
\epsfysize=7 in \epsfbox{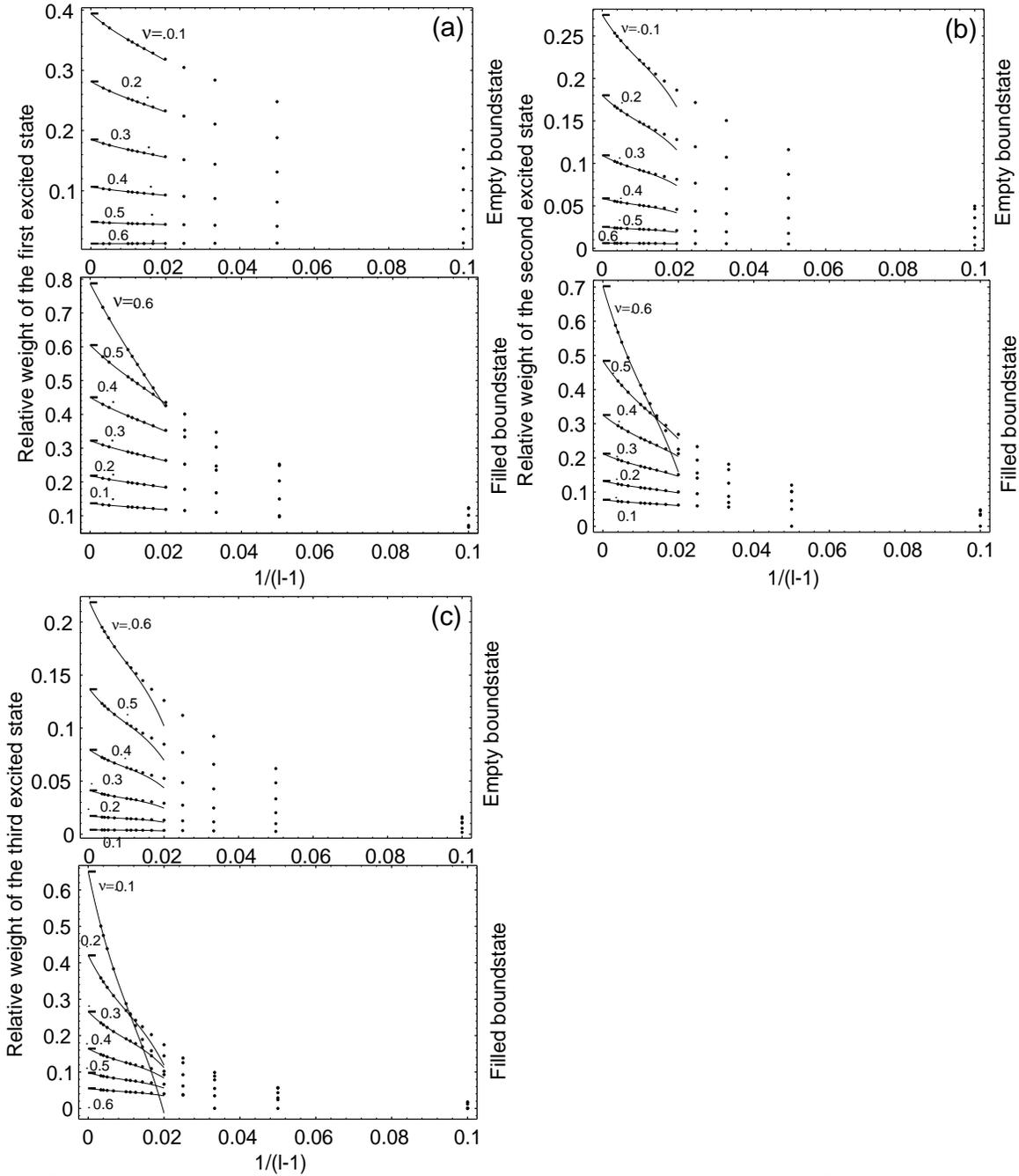}
 \caption{(a)
Relative weight of the first excited state peak in the hole
propagator for different densities $N/(l-1) = 0.1(0.1)0.6$; $t/V = 0.1$. The horizontal lines
mark the  CFT predictions (Eq.(\protect{\ref{peaks}})), curves are the least squares' fits
of the third-order in $1/(l-1)$, calculated from  the four points with smallest
$1/(l-1)$. Notice that these curves are in a good agreement with the 
  data in a larger range of $1/(l-1)$. 
(b),(c) The same for the second and third excited states.} \label{Figure7} \end{figure}

 The calculations of the intensities of 
the first few peaks ( $t/V=0.1$;
Fig.\ref{Figure7})
  shows that finite-size corrections to these amplitudes are
significant  for $l \sim 400$.  Thus we shouldn't expect to be able to
obtain the FES exponent reliably this way even if the neglect of multi
particle-hole excitations was valid.   

In Fig. (\ref{Figure7}) and the Table we show the ratio of the
{\it total} amplitude of all peaks at excitation energy $v_F\pi m/l$,
to the amplitude of the lowest peak ($m=0$) for $m=1,2,3$ for both
filled and empty boundstate, $R^{e,f}_m$, following the discussion in
Sec. II. $t/V=.1$ and the maximum length considered was $l=400$.
The predictions for $R^{e,f}_m$ from conformal field theory
in Eq. (\ref{peaks}) are also shown in the figures and table.  The
finite-size corrections to these ratios are quite large, but upon
extrapolating in $1/l$ we obtain good agreement with 
the CFT
predictions in all six cases.

\noindent  
The finite size corrections,
at a fixed value of $l$,
are roughly
proportional to $\left(\frac{\tilde{\delta}_F}{\pi}\right)^2$, where
the effective phase shift in the presence of bound state  is
$\tilde{\delta}_F  = \delta_F$ if the boundstate is filled, and
$\pi-\delta_F$, if it is empty \cite{CN,Hopfield}. Thus CFT predictions
in a finite system with filled boundstate are more accurate for low 
density than for high density, and the opposite for
empty boundstate.


\section{Sum Rules and Open Questions}

The Fourier transform of the core hole Green's function, proportional
to the photoemission intensity, can be written:
\begin{equation} 
I(\omega ,\nu ) = \sum_n|\left<0|\tilde n\right>|^2\delta (\tilde E_n-E_0-\omega
), \end{equation}
where $\left|\tilde n\right>$, $\tilde E_n$ label all states of the system with
the core hole potential turned on and $\left|0\right>$, $E_0$ refers to the
groundstate without the core hole potential. Here   we make explicit
the fact that $I$ depends on the electron density (i.e. $k_F$). The
states $\left|\tilde n\right>$ can be classified according to whether the
boundstate is filled or empty and accordingly we may decompose
$I(\omega )$: \begin{equation} I(\omega ,\nu )=I_f(\omega
,\nu )+I_e(\omega ,\nu ).\end{equation}

 Near a threshold, $I(\omega ,\nu )$ takes the form:
\begin{equation}
I(\omega ,\nu ) = F(\nu )(\omega-\omega_0)^{-\alpha (\nu
)}.\label{sing}\end{equation}  
So far, we have focussed on the value of the FES exponent, $\alpha
(\nu)$.  In this section we would also like to consider the
dependence of the amplitude factor, $F$ on the density and also the
behaviour of $I(\omega ,\nu )$ away from the threshold. While most of
this behaviour is clearly non-universal, one might expect certain
universal features to immerge in the limit $\nu \to 0$.  Our interest
in this limit is motivated by the experiments on doped
semiconductors.\cite{Young} 

To begin with we  point out
the existence of two sum rules.  These apply very generally to FES
problems in arbitrary dimensions, without any particular assumptions
about spherical symmetry or about the location of the core potential
in  the one-dimensional case. It follows from completeness of the
states $\left|\tilde n\right>$ that $I(\omega )$ obeys the sum rule:
\begin{equation} \int_{-\infty}^\infty d\omega I(\omega ,\nu
)=1.\end{equation}  This implies that the integrated intensities from
the states with filled or empty boundstate obey:
\begin{equation} I_f+I_e=1.
\end{equation}  
Another useful sum rule
can be derived by writing $I_f(\omega ,\nu )$ in terms of the
projection operator, $\hat n_B$ onto states in which the boundstate
is occupied:
\begin{equation} 
\int d\omega I_f(\omega ,\nu )=
\sum_n|\left<0|\hat n_B|\tilde n\right>|^2
=\left<0|\hat n_B|0\right>.\end{equation}  We may write 
\begin{equation} \hat n_B=\psi_B^\dagger \psi_B, \end{equation}
where $\psi_B$ annihilates the boundstate electron.  $\psi_B$ can be
expressed in terms of the operator $\psi_j$ which annihilates an
electron at site $j$ and the boundstate wave-function, $\Psi^B_j$:
\begin{equation} 
\psi_B=\sum_j\Psi^B_j\psi_j.
\end{equation} 
 Thus:
\begin{equation}
I_f=\sum_{i,j}\Psi^B_i\Psi^B_j\left<0|\psi^\dagger_i\psi_j|0\right>.\end{equation}
In a d-dimensional continuum formulation this becomes:
\begin{equation} I_f=\int d^d\vec rd^d\vec r'\Psi^B(\vec
r)\Psi^B(\vec r')\left<0|\psi^\dagger (\vec r)\psi (\vec
r')|0\right>.\end{equation}
At this point, it is convenient to Fourier expand the position-space
annihilation operators.  The form of this expansion depends somewhat
on the boundary conditions (in the system without the core
potential).  In dimension $d>1$ we  generally consider  a
 translationally system in which case:
\begin{equation} \psi_{\vec r} = \int {d^d\vec k\over (2\pi
)^d}e^{i\vec k\cdot \vec r}\psi_{\vec k}.\label{FT}\end{equation}
In this case, the sum rule becomes:
\begin{equation} I_f=\int {d^dk\over (2\pi )^d}|\Psi^B (\vec
k)|^2\theta (\epsilon_F-\epsilon_{\vec k}),\end{equation}
where $\Psi^B(\vec k)$ is the Fourier transform of the boundstate
wavefunction.  The $\vec k$-integral is only over states below the
Fermi surface.  This makes it clear that $I_f$ vanishes as $\nu \to 0$,
and approaches 1 as $\nu \to 1$.  Roughly speaking, when the density
is small, there is a negligible probability of an electron being
near the origin when the core potential is switched on so the overlap
of the unperturbed groundstate with any state in which the boundstate
is occupied goes to 0.  In the opposite limit of high density there
is probability near 1 of an electron being near the origin.  It is
interesting to consider how rapidly $I_f$ vanishes as $\nu \to 0$. 
If we assume that the dispersion relation and core potential are
spherically symmetry, then we may classify the boundstate by its
principle angular momentum quantum number, L.  At small $k$,
$\Psi^B(\vec k )\propto k^L$, so
\begin{equation}I_f\propto k_F^{2L+d}\propto
\nu^{1+2L/d}.\end{equation}

In a one-dimensional translationally invariant system with reflection
symmetry the boundstate can be classified as being an even or odd
function of $x$.  $I_f\propto \nu$ for an even boundstate or $\nu^3$
for an odd boundstate.  When the potential is at the end of a
one-dimensional chain with a free boundary condition,
\begin{equation} \Psi^B(k)\equiv \sum_{j=1}^{l-1}\sin kj \psi^B_j.
\end{equation}  Thus $\Psi^B(k)\propto k$ as $k\to 0$, so $I_f\propto
\nu^3$.

The behaviour of the FES exponent, $\alpha$ at $\nu \to 0$ follows,
in some cases, from Levinson's theorem, which determines the behaviour
of the phase shift as $k\to 0$.  For an s-wave boundstate, or for a
one-dimensional problem with the impurity at the end of the chain,
the phase shift approaches $\pi $ at the bottom of the band when
there is a boundstate.  Thus it follows, from Eqs. (\ref{x_f}),
(\ref{x_e}) and (\ref{alpha}) that $\alpha_f\to 0$ and $\alpha_e\to
1$.  Thus $I_f(\omega )$ becomes a step function near its threshold
and $I_e(\omega )$ becomes a $\delta$-function.  [$\alpha =1$
corresponds to a constant Green's function in the time-domain whose
Fourier transform gives a $\delta$-function.]  Thus $I_e (\omega ,\nu
)$ approaches, in the $\nu \to 0$ limit, the result for the empty
system: a $\delta$-function of unit intensity.  This follows since
the groundstate with no electrons is the same with or without the core
potential; it is simply the vacuum state. Thus we expect 
\begin{equation} F_e(\nu )\to 1,  \ \ (\nu \to 0).\end{equation}
The step in
$I_f(\omega ,\nu )$ corresponds to electrons from the continuum
falling into the boundstate after it is created. The probability for
this process, and hence $F_f(\nu )$ should vanish as $\nu \to 0$.

We have been unable to understand, from general arguments, how
$F_f(\nu )$ approaches 0 or how $F_e(\nu )$ approaches 1 as $\nu \to
0$.  More generally, we would like to know how the functions
$I_e(\omega )$ and $I_f(\omega )$ behave even away from the threshold
as $\nu \to 0$.  In particular, it is interesting to ask whether
there might be some sort of universal scaling form in that limit.
This behaviour is only weakly constrained by the above
sum rules.

\begin{figure}[p] 
 \epsfxsize= 7 in  \epsfbox{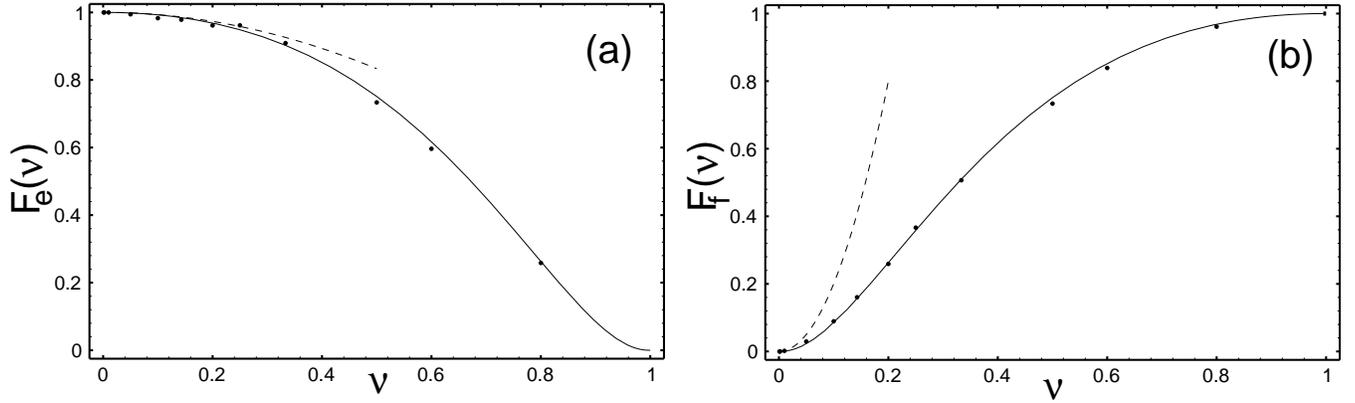}
\caption{FES amplitude as a function of density. The
values of $F_{e}$ (a) and  $F_{f}$ (b)
are plotted vs. $\nu$. The curves are 
$F^{\star}_e(\nu) = (1-\nu)^{2(1-  (1-\nu)^{1/3})}$ and  $F^{\star}_f(\nu) =
\nu^{2(1-\nu^{1/3})}.$ Dotted lines represent
$1-2\nu^2/3$ and  $20 \nu^2$ resp. (see text)} \label{Fig:DensityDep} \end{figure}

\begin{figure}
\epsfxsize= 7 in  \epsfbox{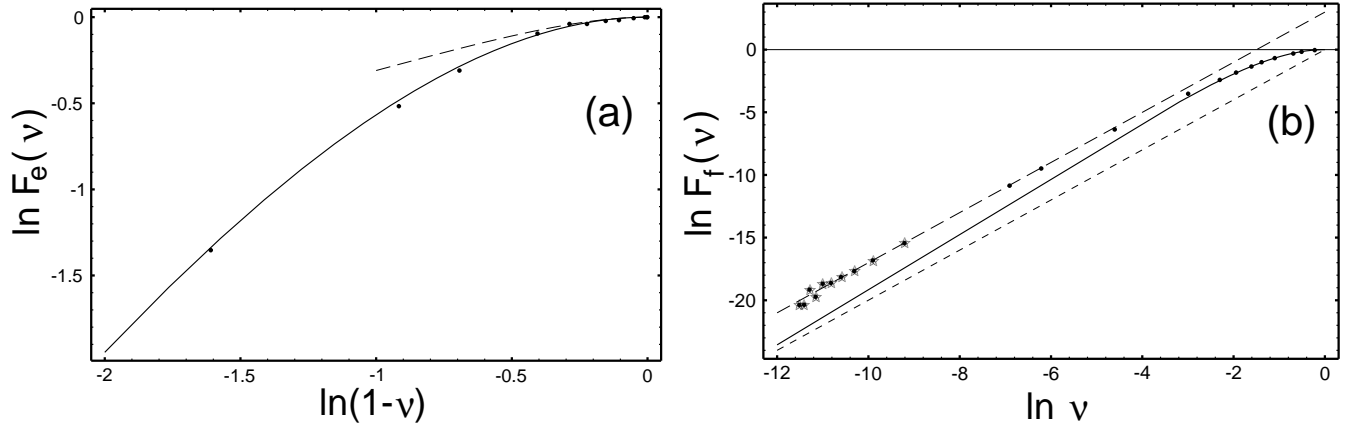}
\caption{FES amplitudes at low densities. In (b) stars represent the data not shown in Fig.6(b); the relation $F_f(\nu) \approx 20 \nu^2$,
the upper dashed line,  is clearly 
seen. The lower dotted line is $\nu^2$, the asymptotics of 
$F^{\star}_f(\nu)$ at $\nu\to 0$.}\label{Fig:DensityDep2}
\end{figure}

It is interesting to investigate these questions numerically. 
For a finite system, $I(\omega ,\nu )$ is a sum of
$\delta$-functions corresponding to the discrete finite size
spectrum. For a large system the intensities of the first few peaks,
with the boundstate filled or empty, will  all be proportional to
$F_{f}(\nu )$ or $F_{e}(\nu )$ respectively.  This follows immediately
from the conformal transformation of Sec. II.  Thus we may conveniently
determined $F_{f,e}(\nu )$ numerically, for a large finite system,
from the groundstate overlap (Anderson orthogonality calculation).  That is:
\begin{equation} |\left<0|\tilde 0\right>|^2=F(\nu) \left({\pi \over
l}\right)^{2x}.\end{equation}
The
resulting functions, $F_{f,e}(\nu )$ are plotted in Fig. (\ref{Fig:DensityDep}) for the
special tight-binding model considered in the previous two sections,
in the limit $t/V\to 0$.  These are obtained from the intercepts of
the curves in  Fig.\ref{Fig:AndersonExp}.  As can be seen
from Fig.\ref{Fig:DensityDep2}, at small $\nu$,


\begin{eqnarray}F_f(\nu )&\to& 20\nu^2\nonumber \\
F_e(\nu )&\to& 1-(2/3)\nu^2. \label{Fapprox}
\end{eqnarray}
The former equation suggests the scaling hypothesis:
\begin{equation} I_f(\omega ,\nu )\to \nu^2 f[(\omega -\omega_0)/\nu
],\label{sh}\end{equation}  for some scaling function $f$.  This is
consistent with the sum rule since:
\begin{equation} \int_{\omega_0}^\infty d\omega I(\omega )
=\nu^3\int_0^\infty dx f(x)\propto \nu^3.\end{equation}
In the $\nu \to 0$ limit, the threshold frequency $\omega_0$
approaches $\epsilon_B-\epsilon_0$, the binding energy
measured from the bottom of the band.  The function $f(x)$ must
approach a constant at the threshold, $x\to 0$, consistent with the
behaviour of the FES exponent $\alpha_f\to 0$ as $\nu \to 0$.  At
large frequencies the function f must vanish sufficiently rapidly for
the integral to converge. Apparently the scale over which $I(\omega )$
varies is set by $\nu \propto v_F$.   If this is
also the relevant energy scale for a translationally invariant
one-dimensional system, or a higher dimensional s-wave boundstate,
then we would expect the behaviour:
\begin{equation} I(\omega ,v_F)\propto f[(\omega -\omega_0)/v_F],
\end{equation} consistent with the sum rule. That is, the threshold 
peak   has a fixed amplitude, but the energy scale over which $I(\omega
)$ decreases scales to 0 as $\nu \to 0$.
[We can speculate that in our model, due to the suppression of 
the wave function near the end
of the chain, the extra factor of $\nu^2$ appears in the expression 
for $I_f$, which would be absent in the $s$-wave channel of a higher-dimension
system, where the sum rule should yield $\nu$ instead of $\nu^3$.]  It would be interesting
to investigate this behaviour in more general models. 

For almost all values of $\nu$ we found that $F_{f,e}$ are well described,
for our special model, by 
the functions  \begin{eqnarray} F^{\star}_f(\nu) =
\nu^{2(1-^3\sqrt{\nu})};\ \ 
 F^{\star}_e(\nu) = (1-\nu)^{2(1-^3\sqrt{1-\nu})}, 
\label{Fstar}
\end{eqnarray}
as shown in Figs.\ref{Fig:DensityDep},\ref{Fig:DensityDep2}. (In two
dimensions the dependence $F_f(\nu) \sim \nu^{2(1-\nu^a)}$ could be
expected from calculations based on the linked cluster approximation
\cite{N}, but with $a=2$, not $1/3$.) At very low densities  $F_e(\nu)$
is still well described by (\ref{Fstar}), as evident from (\ref{Fapprox}) and Fig.\ref{Fig:DensityDep}, while $F_f(\nu)$ approaches the $20 \nu^2$-dependence
of (\ref{Fapprox}) (see  Fig.\ref{Fig:DensityDep2}).

\section{Conclusions}

We have investigated the Anderson orthogonality catastrophe and Fermi edge
singularity in photoemission spectrum in a tractable 1D tight-binding
model of spinless electrons, in the case where the 
core potential produces a boundstate.

We have confirmed the predicted relation between the  scattering phase
on the  Fermi surface, $\tilde{\delta}_F$, and the Anderson and FES
exponents.  We have calculated the ratios of intensities of discrete
adsorption peaks for a finite system, using CFT and checked the formulas at both
primary and secondary thresholds numerically.  We have found that the
higher order finite size corrections are roughly proportional to
$\tilde{\delta}_F^2$ and can be significant in a system as large as several
hundred sites.  Thus they might be observable
  in a mesoscopic system.  The CFT-based relation between the
exponents and $O(1/l)$-term in the ground state energy shift  was
confirmed as well. Using the model, we obtained the explicit density
dependence of the FES amplitude in the whole range of $\nu$.

\acknowledgements

We would like to thank G.~Sawatzky, J.~Young, and R.~Eder for stimulating discussions and for acquainting us with their results prior to
publication. A.Z. is grateful to P.~Hawrylak for interesting discussions and hospitality during his visit to IMS-NRC. This research was supported by NSERC. 
 
\newpage
 \begin{table}[p] \begin{tabular}{|l|l|l|l|l|l|l|l|}     Empty &
$\nu $  & 0.1   & 0.2 &0.3 &0.4 &0.5 &0.6 \\
boundstate & & & & & & & \\
  \hline\hline $m=1$&  & & & & & & \\   &
$(\delta_F/\pi-1)^2$ &  0.0123
 & 0.0485 & 0.1071 &
0.1859 & 0.2827
 & 0.3960 \\ & & & & & & & \\
 & relative weight &  0.0122 & 0.0483 &
0.1067&  0.1851&
0.2816&  0.3944\\ \hline $m=2$&
$\frac{1}{2}(\delta_F/\pi-1)^2 \cdot$ & & & & & & \\ 
  & $ \cdot (1+(\delta_F/\pi-1)^2)$ & 
0.0062 & 0.0254  & 0.0593 &
0.1102 & 0.1813
 & 0.2764 \\ & & & & & & & \\
 & relative weight&  0.0062 & 0.0253 &
0.0589&  0.1095&
0.1802 &  0.2747\\ \hline $m=3$&
$\frac{1}{6}(\delta_F/\pi-1)^2
(1+(\delta_F/\pi-1)^2)\cdot$ & & & & & & \\
  & $\cdot(2+(\delta_F/\pi-1)^2)$
&  0.0042& 0.0174   &
0.0417 & 0.0803 &
0.1380
 & 0.2207 \\ & & & & & & & \\
 & relative weight&  0.0041 & 0.0172 &
0.0412 & 0.0795&
0.1367  &  0.2187\\ \hline\hline
Filled &
$\nu $  & 0.1   & 0.2 &0.3 &0.4 &0.5 &0.6 \\
boundstate & & & & & & & \\
  \hline\hline $m=1$&  & & & & & & \\   &
$(\delta_F/\pi)^2$ &  0.7906
 & 0.6079 & 0.4525 &
0.3236 & 0.2193
 & 0.1374 \\ & & & & & & & \\
 &  relative weight&  0.7868& 0.6051 &
0.4505&  0.3222&
0.2184&  0.1369\\ \hline $m=2$&
$\frac{1}{2}(\delta_F/\pi)^2 \cdot$ & & & & & & \\    & $ \cdot
(1+(\delta_F/\pi)^2)$ & 
 0.7078
 & 0.4887  & 0..3287 &
0.2142& 0.1337
 & 0.0782 \\ & & & & & & & \\
 & relative weight&  0.7025 & 0.4853 &
0.3264& 0.2128&
0.1329 & 0.0777\\ \hline $m=3$&
$\frac{1}{6}(\delta_F/\pi)^2
(1+(\delta_F/\pi)^2)\cdot$ & & & & & & \\
 & $\cdot(2+(\delta_F/\pi)^2)$
&  0.6584 & 0.4248   &
0.2687 & 0.1659 &
0.0989
 & 0.0557 \\ & & & & & & & \\
 & relative weight&  0.6505 & 0.4202 &
0.2659& 0.1642 &
0.0980  &  0.0552\\ \hline
\end{tabular} \caption{Relative weights of first $m$ excited peaks in the
hole propagator compared to CFT predictions of Eq.(2.22). The
relative weights in the limit $l\to\infty$ are obtained from the
best fit to the finite-size  values (Fig.7),
$w(l)=c_0+c_1/(l-1)+c_2/(l-1)^2+c_3/(l-1)^3$ using the method of the least
squares to determine $c_{0,...,4}$ from  the {\em four} points with
smallest
$1/(l-1)$ for each graph.  $w(\infty)=c_0$.}\label{Table} \end{table}

\appendix
\section{Finite size energy}

In this appendix we wish to demonstrate explicitly the formula
relating the finite size correction to the groundstate energy
difference to the phase shift at the Fermi surface.  At the same time
we will expose a subtlety in the definition of the $O(1/l)$ term in
this energy difference. This groundstate energy difference contains a
dominant term of O(1).  This is non-universal, depending on the
ultraviolet cut-off in the Dirac fermion theory.  It must be
subtracted correctly to determine the universal $O(1/l)$ correction.
To be concrete, we consider a tight-binding chain of $l-1$ sites,
$j=1,2,3,... l-1$ with free boundary conditions, N electrons and a
scattering potential, $V_j$, localized near $j=0$.   \begin{equation}
H=-\sum_{j=1}^{l-2}(t\psi_j^\dagger \psi_{j+1}+\hbox{h.c.}
+V_j\psi_j^\dagger \psi_j) \end{equation} When $V_j=0$ the single
particle eigenstates are $\sin kj$.  A simple way of determining the
allowed wave-vectors, $k$, is to imagine adding two ``phantom sites''
at $j=0$ and $j=l$ and then imposing the boundary condition:
\begin{equation}  \psi_0=\psi_{l}=0.\label{phantom} \end{equation} 
 The Fourier expansion of $\psi_j$ in terms of creation and
annihilation operators then involves $\sin (kj)$ with:
\begin{equation} k_n=\pi n/l,\ \ \  m=1,2,3,\ldots
l-1.\label{kn}\end{equation} The groundstate energy for $V=0$ is:
\begin{equation} E_0=\sum_{m=1}^{l-1}\epsilon (k_m),\end{equation} with
$\epsilon (k)=-2t\cos k$.  In fact, this discussion doesn't depend on
the form of $\epsilon (k)$ and can be applied immediately to a more
general Hamiltonian with longer range hopping provided that the
boundary condition of Eq. (\ref{phantom}) applies.   The only property
of $\epsilon (k)$ that we will use is that its derivative vanishes at
$k=0$.

In the $l\to \infty$ limit, the Fermi wave-vector is  \begin{equation}
k_F=\pi \lim_{l\to \infty} N/l.\end{equation} For a finite system,
there is an ambiguity of $O(1/l)$ in the definition of $k_F$ since it
may be chosen anywhere between $\pi N/l$ and $\pi (N+1)/l$.  It turns
out to be convenient to choose it to lie exactly halfway between the
$N^{\hbox{th}}$ and $l^{\hbox{st}}$ level: \begin{equation}
k_F\equiv \pi (N+1/2)/l.\label{kFdef}\end{equation} This gives the
model an approximate particle-hole symmetry, in the vicinity of the
Fermi surface.  [Only at half-filling does this particle-hole symmetry
become exact.]  We regard $k_F$ as being held fixed as $l$ is varied,
for purposes of determining the term of $O(1/l)$ in the groundstate
energy.  Thus the quantity $(N+1/2)/l$ must be held fixed.  In
practice, for numerical simulations, this is not particularly more nor
less difficult than holding fixed the actual density, $N/(l-1)$.  For
instance, to obtain $k_F=\pi /4$ we may choose the number of sites
$l-1=4N+1$ for arbitrary positive integer $N$.

The continuum limit Dirac theory is defined by only keeping
wave-vectors near $\pm k_F$, writing: \begin{equation} \psi_j\approx
e^{-ik_Fj}\psi_L(j)+e^{ik_Fj}\psi_R(j), \end{equation} where
$\psi_{L,R}$ are left and right moving Dirac fields. The boundary
conditions of Eq. (\ref{phantom}) imply: \begin{eqnarray}
\psi_L(0)+\psi_R(0)&=&0 \nonumber \\
 e^{-ik_Fl}\psi_L(l)+e^{ik_Fl}\psi_R(l)&=&0.\end{eqnarray}  Using Eq.
(\ref{kFdef}) the last equation gives: \begin{equation}
\psi_L(l)-\psi_R(l)=0,\end{equation}
 corresponding to the ``same'' boundary conditions at both ends as discussed
in Section II.
$k_F$ was chosen in Eq. (\ref{kFdef}) above in order to obtain this boundary condition on
the Dirac fermion.

Including the scattering potential, $V_j$, the single-particle
eigenstates still become asymptotically plane waves, $\sin [\tilde
kj+\delta (\tilde k)]$ (at distances large compared to the range of
$V$), where $\delta (\tilde k)$ is the phase shift.  Thus the allowed
wave-vectors are now: \begin{equation}  \tilde k_n=k_n-\delta (\tilde
k_n)/l,\label{ps}\end{equation} with $k_n\equiv \pi n/l$.  It is
important to note that the argument of $\delta$ in Eq. (\ref{ps}) is
$\tilde k_n$, not $k_n$.  Thus, to $O(1/l^2)$, we may write:
\begin{equation} \tilde k_n\equiv f(k_n)=k_n-\delta (k_n)/l +\delta
'(k_n)\delta (k_n)/l^2. \label{fdef}\end{equation} The groundstate
energy can thus be written: \begin{equation}
E_0=\sum_{n=1}^{l-1}\epsilon [\tilde k_n].\end{equation} This can be
evaluated using the Euler-MacLaurin expansion: \begin{equation}
\sum_{m=1}^{l-1}F(m-1/2)=\int_0^{l-1}dx F(x)-{1\over 24}[F'(N)-F'(0)]+
O(F'').\label{EM}\end{equation}  Setting: \begin{equation} F(n-1/2) =
\epsilon [f( \pi n/l)],\end{equation}  where the function, $f$ is
given by Eq. (\ref{fdef}), we obtain the convenient result,
\begin{equation} F(N) = \epsilon [f(k_F)].\end{equation} Thus, to
$O(1/l)$: \begin{equation} E_0 = \int_0^{l-1}dn \epsilon \{f[\pi
(n+1/2)/l]\}-v_F\pi /(24l),\end{equation} where \begin{equation}
v_F\equiv \epsilon '(k_F),\end{equation} and corrections of $O(1/l^2)$
have been dropped.  Now it is convenient to change integration
variables to: \begin{equation} k=\pi (n+1/2)/l,\end{equation} giving:
\begin{equation} E_0 = l\int_0^{k_F}{dk\over \pi}\epsilon [f(k)]-{v_F\pi
\over 24l}.\end{equation}  Here the lower limit of integration has been
shifted by $\pi /2l$.  This is justified since $\epsilon (k)$ is
quadratic at $k\to 0$, producing only corrections of $O(1/l^2)$ to
$E_0$.  Using Eq. (\ref{fdef}) and expanding to $O(1/l)$ we obtain:
\begin{equation} E_0= l\int_0^{k_F}{dk\over \pi}\left[ \epsilon
(k)-{\epsilon '(k)\delta (k)\over l}+{\epsilon ''(k)\delta^2(k)\over 2l^2}+{\epsilon
'(k)\delta '(k)\delta (k)\over l^2}\right] -{v_F\pi \over 24l}\end{equation} 
Integrating by parts, and using $\epsilon '(0)=0$, we finally obtain:
\begin{equation} E_0=l\int_0^{k_F}{dk\over \pi} \epsilon (k)-{1\over
\pi}\int_{\epsilon_0}^{\epsilon_F}d\epsilon \delta (\epsilon
)+{\pi v_F\over l}\left\{ {1\over 2}\left[{\delta (k_F)\over
\pi}\right]^2-{1\over 24}\right\}+ O\left({1\over
l^2}\right).\end{equation}  Here $\epsilon_0\equiv \epsilon (0)$ and
$\epsilon_F\equiv \epsilon (k_F)$.  The first term, of $O(l)$ is the
bulk groundstate energy.  The second term of $O(1)$ is a well-known
result referred to as Fumi's theorem.  Note that these terms depend on
$\epsilon $ and $\delta$ over the whole band.  On the other hand, the
final term, of $O(1/l)$ depends only on data right at the Fermi
surface, namely $v_F$ and $\delta (k_F)$.  The $1/24$ term is the
well-known conformal field theory result for open boundary
conditions.  The additional $[\delta (k_F)/\pi]^2$ term gives the
effect on the groundstate energy of changing the boundary conditions.
This formula can be checked for the nearest neighbour model, with
$\epsilon (k)=2t\cos (k)$, and free boundary conditions. The exact
groundstate energy is given by a geometric series: \begin{eqnarray}
E_0&=&t-{t\sin k_F\over \sin (\pi /2l)}\nonumber \\ &=&{l\over
\pi}v_F+t-{\pi v_F\over 24 l}+O(1/l^2),\label{E_0}\end{eqnarray} with
$v_F\equiv 2t\sin k_F$.

 We  note that adopting a different definition of $k_F$, such as $\pi
N/(l-1)$ in the one dimensional model, corresponds to adding a 
$k$-independent term to the phase shift [$(\nu-1/2)$, where $\nu$ is
the density, in this case].  Eq. (\ref{E_0}) still holds, when written
in terms of this redefined phase shift.  This discussion was given for
the case of a one dimensional tight-binding model with free ends but
it can be easily generalized, for example to the s-wave sector of a
three dimensional spherically symmetric continuum model with a
vanishing boundary condition on the surface of a sphere if radius,
$l$.  With an appropriate definition of $k_F$ [essentially defining,
to $O(1/l)$ what is held fixed as $l$ is varied], Eq. (\ref{E_0}) is
again obtained. 

We can easily generalize Eq. (\ref{E_0}) to calculate the energy of a
state with $n$ extra electrons added, with $n$ held fixed as $N$ and
$l\to \infty$.  This gives: \begin{eqnarray}
E_n&=&E_0+\sum_{m=1}^n\epsilon [k_F-\delta (k_F)/l+\pi
(m-1/2)/l]\nonumber \\ &=&E_0+n\epsilon_F+(v_F\pi
/l)\sum_{m=1}^n(m-1/2-\delta (k_F)/\pi)+O(1/l^2).\end{eqnarray} Hence
\begin{equation}E_n=l\int_0^{k_F}{dk\over \pi} \epsilon (k)-{1\over
\pi}\int_{\epsilon_0}^{\epsilon_F}d\epsilon \delta (\epsilon
)+n\epsilon_F+ {\pi v_F\over l}\left\{ {1\over 2}\left[n-{\delta
(k_F)\over \pi}\right]^2-{1\over 24}\right\}+ O\left({1\over
l^2}\right).\end{equation}

 \section{Dispersion law and wave functions in a 1D tight-binding chain} 

Here we calculate exactly the phase shift, finite size spectrum,
eigenstates and overlaps for the one-dimensional tight-binding model with
an impurity potential at one end, given in Eq. (\ref{Hsolv}).

It can be easily seen that the eigenstates can be written exactly in the
form: \begin{equation} \Psi_j \propto \sin k(j-l).\label{wf}\end{equation}
This wavefunction trivially satisfies the lattice Schroedinger equation
(for arbitrary $k$) at all sites 2,3, $\ldots$, $l$-1, with:
\begin{equation} \epsilon (k) =-2t \cos k.\label{eigen}\end{equation} The
Schroedinger equation for the first site determines the allowed values of
$k$: \begin{equation} -t\Psi_2-V\Psi_1=\epsilon \Psi_1.\end{equation}
Inserting Eq. (\ref{wf}) and (\ref{eigen}), we obtain: \begin{equation}
{\sin k(l-1)\over \sin kl}={t\over V}.\label{detk} \end{equation} For
sufficiently large $V/t$ there is also a boundstate with wavefunction:
\begin{equation} \chi_j\propto \sinh \kappa (l-j),\end{equation} where
$\kappa >0$ in order that the wavefunction decreases with increasing $j$.
This has energy: \begin{equation} \epsilon_B=-2t\cosh \kappa
.\end{equation} Again the Schroedinger equation is satisfied
automatically (for any $\kappa$) at all sites except the first which
gives the condition determining $\kappa$: \begin{equation} {\sinh \kappa
(l-1)\over \sinh \kappa l}={t\over V}.\label{kappa}
\end{equation}  For $l>>1$, this gives:
\begin{equation} e^{-\kappa}=t/V.\end{equation}
Since $\kappa$ must be positive, we see that there is only a boundstate
solution for $V>t$.  In this case:
\begin {equation} \epsilon_B=-(V+t^2/V).\end{equation}
 For the continuum states we may
calculate the exact phase shift, defined by the form of the wavefunction
for $l>>1$: \begin{equation} \Psi_j\propto \sin [kj+\delta
(k)].\end{equation} From Eq. (\ref{wf}) we see that: \begin{equation} -kl
= \delta (k)+ \pi n,\end{equation} for some integer $n$, in the limit
$\l\to \infty$.  Substituting this into Eq. (\ref{detk}), we
obtain:
\begin{equation} {\sin [\delta (k)+k]\over \sin [\delta (k)]}={t\over
V}.\end{equation} This gives:
\begin{equation}\delta = \arctan\left[{\sin k\over t/V-\cos k}
\right] .\end{equation}

We now consider in more detail the spectrum of the finite system, of $l-1$
sites. With no potential, there are $l-1$ band wave-functions with:
\begin{equation} k_n=n\pi /l,\ \  n=1,2,3,\ldots (l-1).\end{equation}
Including the attractive potential, we see from Eq. (\ref{kappa}) that 
there is a solution of the form $\sinh \kappa (j-l)$ for $t/V<1-1/l$. We
label
this solution $\tilde \epsilon_1$ and $\tilde \Psi^1$.  For this range
of $t/V$ Eq. (\ref{detk}) has only $l-2$ solutions, $\tilde k_2$,
$\tilde k_3$, $\ldots$ $\tilde k_{l-1}$. In particular, for $V/t\to
\infty$, these $l-2$ solutions become: $\tilde k_n=(n-1)\pi /(l-1)$,
corresponding to a chain with  free boundary condition at both ends and
$l-2$ sites.  

The normalization of the band states can be calculated exactly in terms of
$\tilde k$ using: \begin{equation} \sum_{j=1}^{l-1}\sin^2k(j-l)={1\over
2}\left[(l-1)- {\sin k(l-1)\cos kl\over \sin k}\right].\end{equation} 
Similarly the  normalization of the boundstate is determined by:
\begin{equation} \sum_{j=1}^{l-1}\sinh^2\kappa (j-l)={1\over
2}\left[(l-1)- {\sinh \kappa (l-1)\cosh \kappa l\over \sinh \kappa}\right]
.\end{equation} 

The overlaps of band wavefunctions with and without the potential can be
calculated similarly using:
\begin{equation} \sum_{j=1}^{l-1}\sin k(j-l)\sin \tilde k(j-l)=
{\sin k(l-1)\sin l\tilde k - \sin lk \sin (l-1)\tilde k\over
2[\cos \tilde k-\cos k]}.\end{equation} If $k$ and $\tilde k$ are
allowed wave-vectors corresponding to potentials
$V_1$ and $V_2$ respectively, then using Eq. (\ref{detk}), 
\begin{equation} \sum_{j=1}^{l-1}\sin k(j-l)\sin \tilde k(j-l)=
{(V_2-V_1)\sin k(l-1)\sin \tilde k(l-1)\over 2t[\cos \tilde k-\cos k]}.
\end{equation}  Thus we obtain the extremely useful result:
\begin{equation} <\tilde \Psi |\Psi >={(V_2-V_1)C(\tilde k)C(k)\over
\epsilon (k)-\epsilon (\tilde k)},\end{equation} where $\epsilon
(k)=-2t\cos k$ is the band energy and
 \begin{equation} C(k)\equiv
{\sqrt{2} \sin k(l-1)\over \sqrt{(l-1)-\sin k(l-1)\cos kl/\sin
k}}.\end{equation} The corresponding
result involving the boundstate follow immediately upon replacing
$\tilde k$ by $i\kappa$.  $\epsilon (\tilde k)$ simply gets
replaced by $\epsilon_B\equiv -2t\cosh \kappa$ and $C(\tilde k)$  by 
\begin{equation} C_B \equiv {\sqrt{2} \sinh \kappa (l-1)\over
\sqrt{\sinh \kappa (l-1)\cosh \kappa l/\sinh
\kappa-(l-1)}}.\end{equation}
  To calculate the overlap of the $V=0$ groundstate with an arbitrary
state with $V\neq 0$, we simply set $V_1=0$ and $V_2=V$ in the above
formula.  This remarkably simply form for the overlaps of single
particle wavefunctions leads to enormous simplification in the
calculation of the overlap of the Bloch determinant multi-particle
states, as shown in Section IV.

%

\end{document}